\documentclass[aps,prd,twocolumn,amsmath,amssymb,amsfonts,superscriptaddress,floatfix,nofootinbib,longbibliography]{revtex4-2}
\usepackage{epsfig} \usepackage{graphicx}
\usepackage[pdftex,bookmarks]{hyperref}
\usepackage[pdftex]{color}
\usepackage[table]{xcolor}  %table package
\usepackage{graphicx}% per tabella
\usepackage{wasysym} %per il simbolo della luna. Quello della Terra è sbagliato

\usepackage{float} % serve per posizionare le figure nel posto preciso con la sintassi  [H]
 \usepackage[normalem]{ulem}
 \useunder{\uline}{\ul}{}

\newcommand{\bgar}{\begin{eqnarray}}
\newcommand{\enar}{\end{eqnarray}} 
 
 \newcommand{\be}{\begin{equation}}
\newcommand{\ee}{\end{equation}}  
 \def\mincirc{\lower3pt\hbox{$\buildrel<\over{\hbox{$\mathchar"218$}}$}}

\DeclareMathSymbol{\mdot}{\mathord}{symbols}{"01}

\begin{document}

\title{\bf Ground measurements of the gravitational redshift questioned:\\ re-establishing the physical bases\\}

 \date{\today}

\author{Anna M. Nobili}
\affiliation{
Dept. of Physics ``E. Fermi'', University of Pisa, Largo B. Pontecorvo 3,
56127 Pisa, Italy}
\author{Alberto  Anselmi}
\affiliation{Thales Alenia Space Italia, Strada Antica di Collegno 253, 10146 Torino, Italy (ret.)}

\begin{abstract}
\noindent In their paper ``\textit{Matter waves and clocks do not observe uniform gravitational fields}'' Asenbaum, Overstreet and Kasevich \href{https://doi.org/10.1088/1402-4896/ad340c}{(2024)}, motivated by alleged inconsistencies in the scientific and educational literature,  aim to clarify some fundamental concepts in the physics of gravitation. In particular, they reexamine the first experimental measurement of the gravitational redshift by Pound and Rebka in 1960, claiming  that it did not in fact measure the gravitational redshift predicted by Einstein almost half a century earlier,  but rather a Doppler shift originating from non-gravitational reaction forces. We show that their conclusion arises from a misunderstanding of the reference systems involved, along with an unphysical interpretation of non-gravitational forces. Thus, our work restores the Pound and Rebka experiment to its rightful place in the history of Physics. Beyond the specific paper addressed, this analysis re-establishes the physical bases of simple, yet fundamental issues of gravitational physics.\footnote{This paper has been published on \textit{Physica Scripta} under the title: Comment on \textit{`Matter waves and clocks do not observe uniform gravitational fields'}  \href{https://doi.org/10.1088/1402-4896/aeb1fd}{(Nobili and Anselmi, 2026)}, along with a rebuttal  by the authors of the paper under discussion \href{https://doi.org/10.1088/1402-4896/aeb33a}{(Asenbaum, Overstreet and Kasevich,  2026)}. A consideration on the rebuttal can be found here at the end of the Conclusions.}

\end{abstract}

\maketitle

\section{Introduction}
\label{Sec:Introduction}

\noindent Inside a non rotating ``Einstein Elevator'' (EE) freely falling in a flat non rotating Earth in which special relativity rules, the equivalence between inertial and gravitational mass holds for all matter, there are no non-gravitational forces and  all Test Masses (TMs) have zero initial velocity, it is found that gravity has no effects and cannot be detected. With the same assumptions, in the ``Flat Earth'' (FE) system in which the EE is falling it is possible to test the Weak Equivalence Principle (WEP) through the Universality of Free Fall (UFF) and to measure the gravitational redshift as predicted by Einstein in 1911\,\cite{Einstein1911} (available in\,\cite{PrincipleRelativity} with the title ``\textit{On the influence of gravitation on the propagation of light}'') . 

Asenbaum, Overstreet  and  Kasevich\,\cite{AOK2024} (AOK hereafter) express their concerns that at research and education level there are inconsistencies about these fundamental concepts and concentrate on clarifying the underlying issues, in particular regarding the gravitational redshift. 
 
They consider a light source, a detector and an observer freely falling as in the EE system and conclude that the frequency shift is zero --as it should be in this system-- but only because the  Doppler and the gravitational frequency shifts have opposite sign and cancel each other out. Then, by building on this result, they revisit the first experimental measurement of the gravitational redshift by Pound and Rebka in 1960\,\cite{PoundRebka1960} coming to the surprising conclusion --which has motivated this work-- that they did not measure the gravitational redshift but rather a Doppler shift originated from the non-gravitational reaction forces which keep at rest in the lab the emitter and the absorber of $\gamma$ rays used in the experiment, by counteracting their weight. 

Both these conclusions arise from a basic misunderstanding on the reference systems involved in the experiments and the physics therein. In Sec.\,\ref{Sec:Figure2aEE} we show that in the EE system there is neither a gravitational nor an equal and opposite Doppler shift, neither for photons nor for clocks. The result by AOK of a Doppler and a gravitational shift cancelling each other out is correct for a source and a detector falling free in the FE  reference system and observed therein, an obvious result that has nothing to do with what happens inside a free falling Einstein Elevator (within the  assumptions made at the beginning).

In Sec.\,\ref{Sec:PoundRebkaIsCorrect} we analyze, along the same lines, the Pound and Rebka experiment\,\cite{PoundRebka1960} and its re-interpretation by AOK. We show beyond question that such re-interpretation is based on their previous incorrect result plus a new, unphysical statement about the r\^ole of the reaction force which keeps source and detector at rest in the lab. Therefore,  Pound and Rebka have  indeed measured  the gravitational redshift predicted in 1911. 

In concluding  their analysis AOK write that a new measurement of the gravitational redshift might be possible in the near future in which, unlike the Pound and Rebka experiment, no non-gravitational support force is involved to keep the source and the detector at rest, letting them free to fall. We anticipate that, should they be able to carry out such an experiment they must be prepared to finding a net zero frequency shift, even to second order (as long as gravity gradients can be neglected).

In Sec.\,\ref{Sec:Conclusions} we conclude by summarizing the main results of our analysis while adding a few general considerations on the gravitational redshift, the experiments to measure it and their overall physical relevance as compared to tests of the WEP and UFF. 

\section{A basic analysis of gravitational and Doppler frequency shift in  free fall}\label{Sec:Figure2aEE}

\noindent Let us consider a reference system, named ``Flat Earth'' (FE), 
for which the following set of assumptions H holds: 

\begin{itemize}
\item[H1]  FE is at rest on a flat, non rotating Earth and there is no other source mass. The Earth's radius is infinite compared to any distance of physical interest in the system, i.e. there are no gravity gradients (no tidal effects). The limitation to a perfectly uniform gravitational field is set quantitatively  by the ratio of the tidal acceleration between the masses involved  to the local gravitational acceleration from the Earth.  Special relativity rules.
\item[H2] The equivalence between inertial and gravitational mass holds for all matter (clocks and ordinary matter alike), i.e. WEP and  UFF are not violated. The assumption is valid to the level established by the best experimental tests of WEP/UFF of the time.

\item[H3] There are no non-gravitational forces.
\item[H4] All TMs have initial zero velocity.
\end{itemize}

\begin{figure}[H]
\begin{center}
\includegraphics[width=0.22\textwidth]{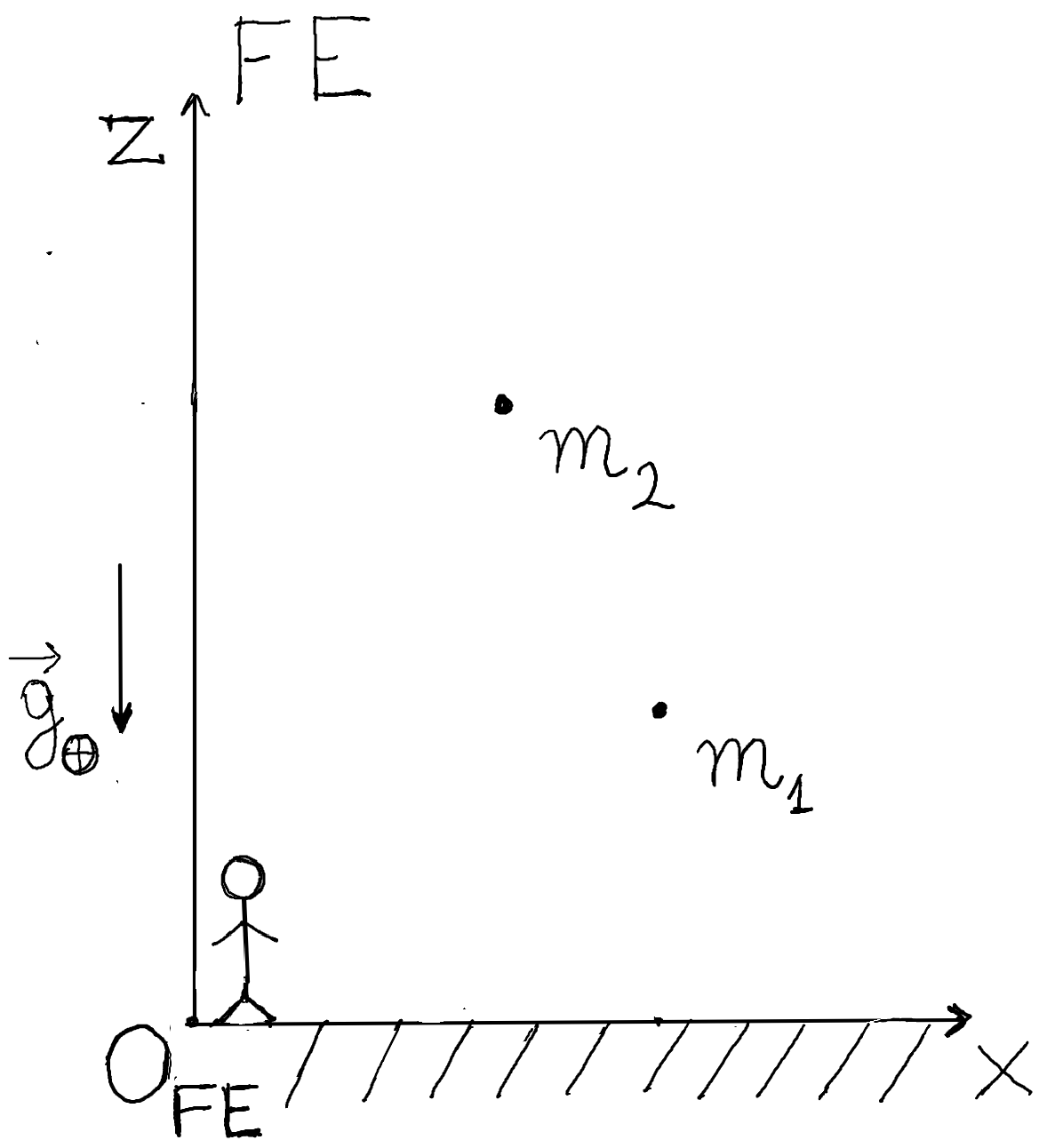}
\caption{2D sketch of the ``Flat Earth'' (FE) inertial reference system in which the set of assumptions H holds and  $\vec g_\oplus$ is the same everywhere ($g_\oplus\simeq9.8\,\rm ms^{-2})$. The free  to fall (unsupported) test masses $m_1,m_2$ are shown at initial time with zero velocity.}
\label{Fig:FEsystem}
\end{center}
\end{figure}

\noindent FE, sketched in 2D in  Fig.\,\ref{Fig:FEsystem}, is an inertial reference system where the same  gravitational acceleration $\vec g_\oplus$ acts at any point (regardless of the presence of a TM or not) because of H1. 

For the free TMs $m_1,m_2$, under the assumptions H, the Observer  $\rm{O_{FE}}$ writes the equation of motion:
\begin{equation}\label{Eq:FEeq}
\ddot z=- g_\oplus
\end{equation}
whereby they both fall along $z$ towards $z=0$ with the same acceleration. During the time of fall they maintain the same relative position as at initial time with increasing but equal velocities.

\begin{figure}[H]
\begin{center}
\includegraphics[width=0.25\textwidth]{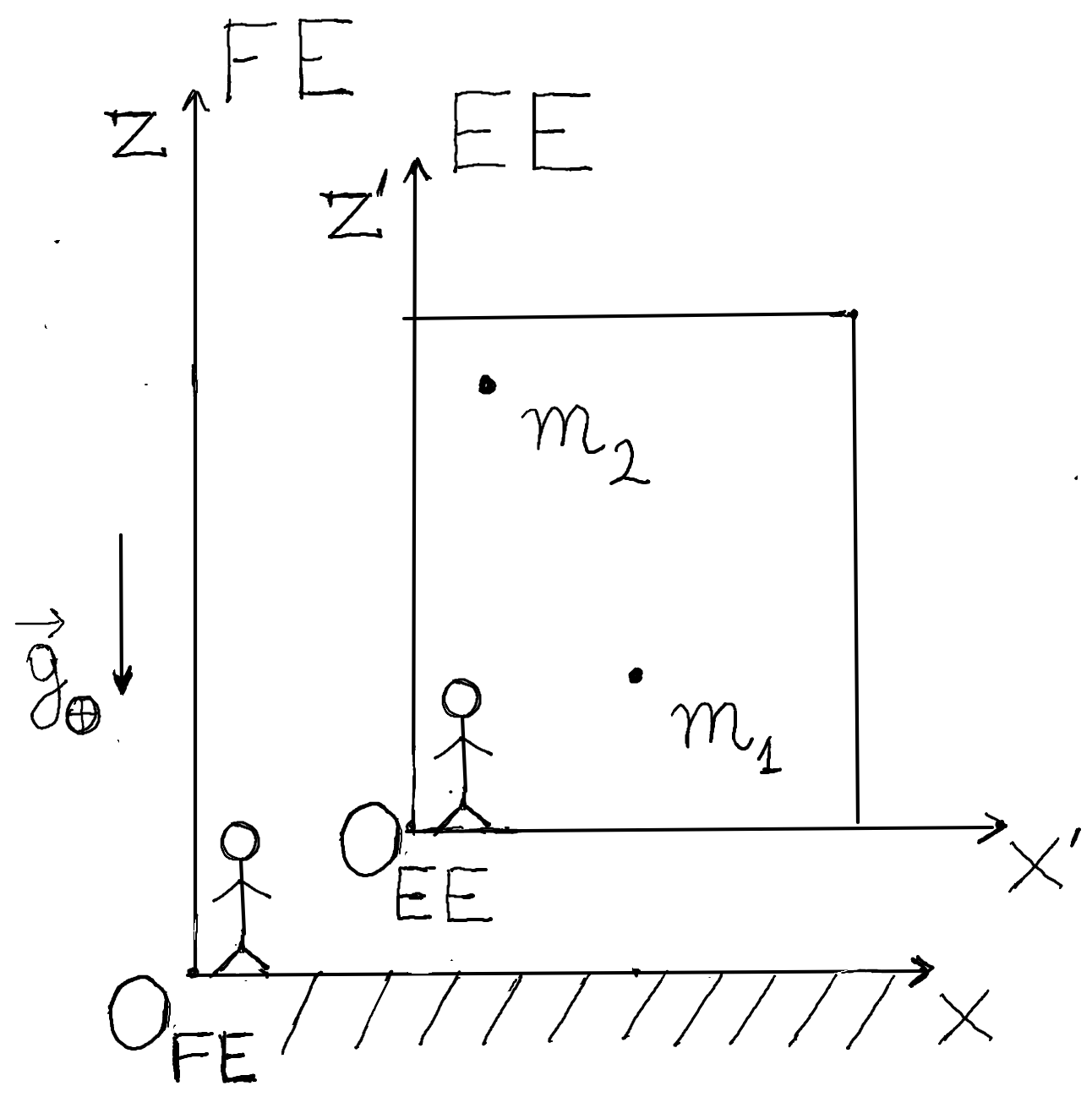}
\caption{The ``Einstein Elevator'' (EE) reference system, enclosing the TMs $m_1,m_2$ and the Observer $\rm{O_{EE}}$, is freely falling in the inertial FE system. EE is an accelerated, non inertial reference system with $x',z'$ the horizontal and vertical axes parallel to $x,z$ respectively.}
\label{Fig:FEwithEE}
\end{center}
\end{figure}

Let us now imagine that the TMs $m_1,m_2$ are enclosed in a (non rotating) ``Einstein Elevator'' (EE) which, with assumptions H, is falling with the same acceleration as the TMs and the Observer $\rm{O_{EE}}$ inside it.  Fig.\,\ref{Fig:FEwithEE} shows both the FE  and the  EE reference systems.

We analyze the physics inside the EE from the point of view of the Observers $\rm{O_{FE}}$ and $\rm{O_{EE}}$, and make sure that they reach the same result.

Even if unable to look inside the EE, the Observer $\rm{O_{FE}}$ can tell that the Observer $\rm{O_{EE}}$ and the TMs remain at rest inside the Elevator  for the entire duration of the fall, by reasoning as follows.
 $\rm{O_{FE}}$ knows that the EE  is an accelerated non inertial system with acceleration $\vec g_{_\oplus}$. Therefore, at any point inside the Elevator there must be an inertial force (proportional to the inertial mass $ m^i $) equal and opposite to the gravitational one (proportional to the gravitational mass $ m^g $). He concludes that the equation of motion inside EE is:
\begin{equation}\label{Eq:FEeqElevator}
m^i\ddot z'=- m^gg_\oplus +m^ig_\oplus
\end{equation}
hence, with H2 ($ m^i=m^g $):
\begin{equation}\label{Eq:FEeqElevator2}
\ddot z'=- g_{_\oplus} +g_{_\oplus}=0 \ \ 
\end{equation}
and therefore the TMs remain at rest  inside EE.

As for  $\rm{O_{EE}}$, he observes that all masses originally at rest remain at rest anywhere in the lab with no need to be supported. This means that EE is an inertial reference frame where the equation of motion for all TMs is: 
\begin{equation}\label{Eq:EEeq}
\ddot z'=0
\end{equation}
as concluded by $\rm{O_{FE}}$ with (\ref{Eq:FEeqElevator2}).

However, should $\rm{O_{EE}}$ be able to look through the $z'=0$ plane of the Elevator, he would discover that the Elevator, himself and the TMs inside it are all falling with the same acceleration, parallel to $z'$, towards the $z=0$ plane. $\rm{O_{EE}}$  can measure this  acceleration, verify that it is  $g_\oplus\simeq9.8\,\rm ms^{-2}$, and conclude --in agreement with $\rm{O_{FE}}$-- that EE is an accelerated reference system where  the correct equation of motion for any test mass is  (\ref{Eq:FEeqElevator2}).

Either way,  (\ref{Eq:FEeqElevator2}) and (\ref{Eq:EEeq}) show that in the free falling EE system  there is no net acceleration. Therefore, an Einstein Elevator freely falling in a uniform gravitational field is equivalent  to an identical Elevator free floating in empty space, i.e. far away from any source mass which might produce a measurable effect. This reference system, named S and shown in Fig.\,3, is inertial, with no force  acting on the TMs. Hence, the observer $\rm{O_{S}}$ writes the equation of motion: 
\begin{equation}\label{Eq:Seq}
\ddot z''=0
\end{equation}
(and the same for the other coordinates) just like Eq.(\ref{Eq:FEeqElevator2}) written by $\rm{O_{EE}}$.
\begin{figure}[H]
\begin{center}
\includegraphics[width=0.25\textwidth]{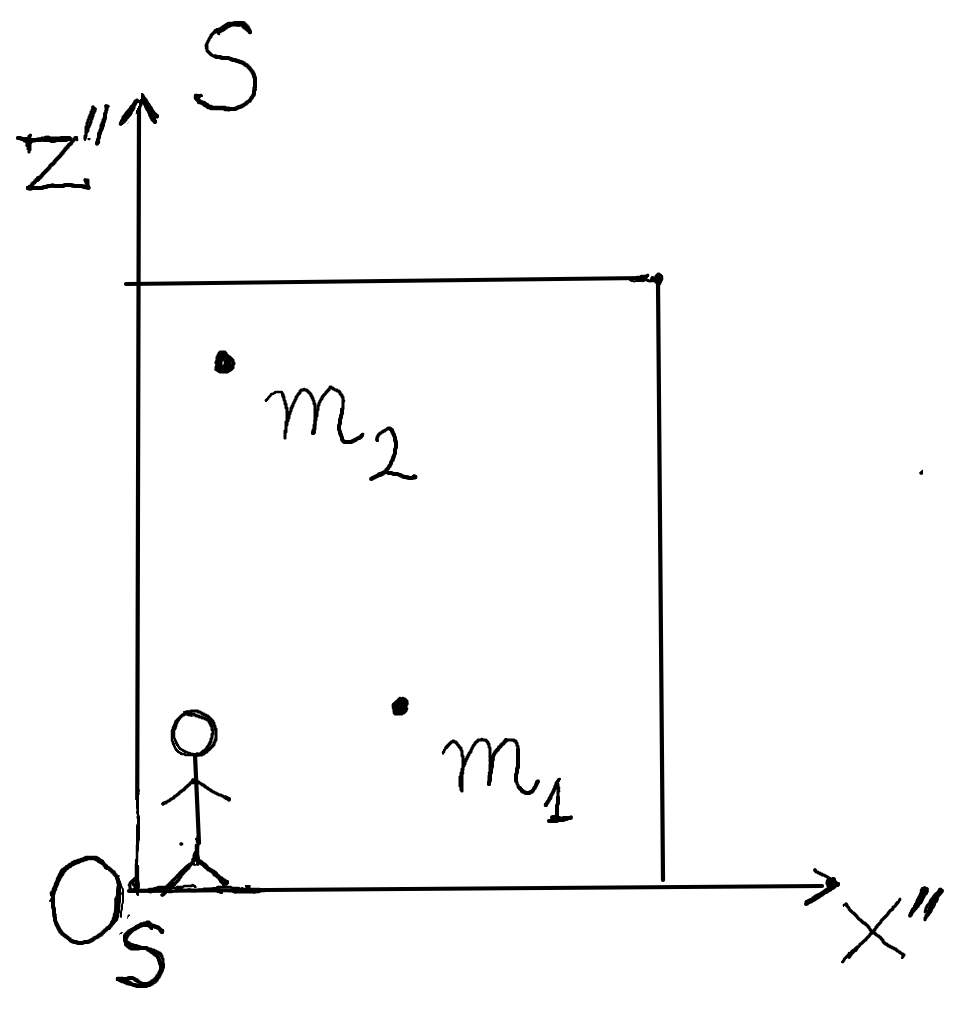}
\caption{The reference system S in empty space with Observer  $\rm{O_S}$.  It is an inertial reference system with no force field, where the   test masses  $m_1,m_2$, with initial zero velocity, are at rest needing no support. It is equivalent to the free falling EE  system shown  in Fig.\,\ref{Fig:FEwithEE}, the axes $x'',x'$ and $z'', z'$ being parallel to each other.}
\label{Fig:S}
\end{center}
\end{figure}
In 1907 Einstein noticed\,\cite{Einstein1907} (available in\,\cite{Schwartz1977} with the title ``\textit{On the relativity principle and conclusions drawn from it}") that in a free falling EE system, because of the equivalence between inertial and gravitational mass (and with all assumptions H) gravity has no effects, and therefore it is undetectable, as it is apparent from (\ref{Eq:EEeq}).

In 1911 he showed\,\cite{Einstein1911} that, in the presence of a uniform gravity field as in the Flat Earth reference system of Fig.\,\ref{Fig:FEsystem} (and assumptions H) photons emitted with energy $E_1$ from a given (fixed)  position and received at a higher one (also fixed) separated by a distance $h$ along the lines of force of the local gravitational field, will have a lower energy $E_2$ at detection than they had at emission, given by:%\footnote{In\,\cite{Einstein1911} it is stressed that this is valid to a first approximation} 
\begin{equation}\label{Eq:PhotonsWeight}
E_2\simeq E_1\Big(1-\frac{g_{_\oplus} h}{c^2}\Big)
\end{equation}
(with $c\simeq3\cdot10^8\,\rm ms^{-1}$ the speed of light which, within special relativity, is the same in all frames), and a corresponding redshift of their respective frequencies\footnote{In\,\cite{Einstein1911} it is stressed that this is valid to a first approximation.}$^,$\footnote{Einstein\,\cite{Einstein1911} obtaines this result by making use of the concepts of gravitational mass and gravitational potential energy of the photon. This has been rightly criticized\,\cite{Russi2020} since the photon has zero rest mass and  it cannot be treated in a non-relativistic manner. However, all that is needed for obtaining the gravitational frequency shift is special relativity and the weak equivalence principle as shown by Schiff in 1960\,\cite{Schiff1960}. Einstein's  formula (\ref{Eq:PhotonsWeight}) holds, but it is the energy difference between two atomic levels that increases with the distance of the atom from the source mass, while the energy of the propagating photon does not change. With this premise, we will continue to use Einstein's old formulation, to facilitate a parallel discussion of photons and clocks.}:
\begin{equation}\label{Eq:PhotonsRedshift}
\nu_2\simeq\nu_1\Big(1-\frac{g_{_\oplus} h}{c^2}\Big)\ \ \ .
\end{equation}

Einstein derives this result in a reference system where the uniform gravitational field $\vec g_{\oplus}$ of the FE system is replaced by an accelerated frame rigid with a ``spaceship'' equipped with a rocket which can accelerate it at a constant acceleration $-\vec g_{\oplus}$ in an otherwise empty space. 
In Fig.\,\ref{Fig:PRandAS} of Sec.\,\ref{Sec:PoundRebkaIsCorrect} we show that this system --that we name Accelerated Spaceship (AS)-- is equivalent (with assumptions H) to the    lab system  of the Pound and Rebka  experiment, that we name PR,  and allows Einstein to establish the frequency shift of light using only special relativity  and the principle of equivalence. He has the source and the detector rigidly attached to the AS system, with separation distance $h$ along the direction of the acceleration, and the process of light emission and detection is investigated in a fixed inertial reference system, which he names $K_\circ$, where there is no force, coinciding with AS at the time of light emission. He adds that all measurements must be made in $K_\circ$ by means of instruments which are proven to give identical results when operated at the same location.

In 1960 Schiff\cite{Schiff1960} showed that the effect of gravitational shift between two clocks $C_1,C_2$ rigidly attached to the PR system of Fig.\,\ref{Fig:PRandAS} (\textit{a}) at --respectively-- lower and higher altitude, can also be obtained with the clocks accelerated in the equivalent (by WEP) AS system of Fig.\,\ref{Fig:PRandAS}(\textit{b}) and observed in the $K_\circ$ fixed system as defined by Einstein. Let us call $z_\circ$ its axis parallel to the $z'''$ axis of the AS system, also parallel to $-\vec g_{_\oplus}$ provided by the rocket.

The Observer $\rm{O_{K_\circ}}$ has 3 identical clocks $C_\circ,C_1,C_2$ which, at the same location in $K_\circ$,  are demonstrated to tick with the same period $\tau_\circ$. They all have zero velocity, $K_\circ$ is inertial and there is no force in it, therefore they don't need to be supported in order to be at rest. Then, in $K_\circ$ the clock $C_\circ$ is moved to a higher altitude along $z_\circ$ where it remains at rest (needing no support) while continuing to tick at $\tau_\circ$. In the meantime the rocket is still off, and the clock $C_2$ is moved a distance $h$ above $C_1$ along $z'''\|z_\circ$   where it is left with zero velocity and so that both $C_1$ and $C_2$ are at lower altitude than  $C_\circ$ while also ticking at $\tau_\circ$\footnote{Note that the clocks $C_1$ and $C_2$, when placed a distance $h$ apart,  are rigidly attached to the spaceship (the AS system of Fig.\,\ref{Fig:PRandAS}(\textit{b}))  in order to keep them at rest relative to it once it moves with acceleration $-\vec g_{\oplus}$, while  as long as the rocket is off, having zero velocity, they are at rest anyway.}.

Once the rocket is turned on and the AS system moves with acceleration $-\vec g_{\oplus}$, the higher clock $C_2$ passes by $C_\circ$ (which is at rest) with velocity $v_2$ and, according to time dilation of special relativity, it  ticks  with a longer period $\tau_2=\tau_\circ(1-v_2^2/c^2)^{-1/2}$ relative to it. Second comes $C_1$, which passes by $C_\circ$ at higher velocity $v_1$, as $v_1^2=v_2^2+2g_{\oplus}h$ and therefore ticks with an even longer period $\tau_1=\tau_\circ(1-v_1^2/c^2)^{-1/2}$. As a result we have:
\begin{equation}\label{Eq:ClocksBlueshift}
\tau_2\simeq\tau_1\Big(1-\frac{g_\oplus h}{c^2}\Big)
\end{equation}
whereby the higher clock (farther away from the source body in the equivalent PR system of Fig.\,\ref{Fig:PRandAS}(\textit{a})) will have a smaller ticking period than a clock located deeper in the field, similarly to what happens to the energy of photons emitted at lower altitude and absorbed at a higher one,  as shown by (\ref{Eq:PhotonsWeight}).

Instead, for the frequency of the clocks $\nu_{_C}=1/\tau$ it is:
\begin{equation}\label{Eq:ClocksFrequencyshift}
\nu_{_C{_{_{2}}}}\simeq\nu_{_C{_{_{1}}}}\Big(1+\frac{g_\oplus h}{c^2}\Big)
\end{equation}
i.e. the sign of the gravitational frequency shift for clocks is opposite to that for photons as given by (\ref{Eq:PhotonsRedshift}).

In the Introduction of Ref.\cite{AOK2024}  AOK assume the equivalence of inertial and gravitational mass and define a ``\textit{Uniform Gravitational Field (UGF)}'' as a small enough region in which the gravitational field is approximately uniform with acceleration $\vec a_{_G}$. They write: ``I\textit{n this UGF every object and observer will fall with acceleration $\vec a_{_G}$. Relative accelerations between objects vanish. A UGF cannot create relative dynamics between observers and test objects because it acts in a universal manner.}''

They are evidently referring to the Einstein Elevator reference system of  Fig.\,\ref{Fig:FEwithEE} (under assumptions H). This is confirmed in Figure\,2(a) of Ref.\cite{AOK2024}, where they show a light source, a detector and the  Observer holding it all freely falling in a uniform gravitational field with acceleration  $\vec{a}_{_G}$.
The light emitted by the source at time $t_{_E}$ reaches the detector at time $t_{_D}=t_{_E}+h/c$ ($h$ being their separation distance along the lines of force of the field). AOK state that by the time $t_{_D}$ the free falling detector has acquired the velocity ${a}_{_G}h/c$ relative to the source (directed towards the source because in their sketch the signal travels towards the detector against the acceleration field) resulting in a Doppler blueshift of the relative frequency by:
\begin{equation}\label{Eq:AOKdoppler}
\frac{\Delta\nu_{_D}}{\nu}\simeq+\frac{{a}_{_G}h}{c^2}
\end{equation}

This effect caused by $a_{_G}$ is  unexpected by AOK and  in contradiction with their  premise that  gravity should be undetectable in this experiment. The way out they find  from such a paradoxical result is that the Doppler blueshift (\ref{Eq:AOKdoppler}) is compensated exactly by an equal and opposite gravitational redshift $\Delta \nu_{grav}/\nu\simeq-a_{_G}h/c^2$ which they argue is there in the same reference system for the same source and detector. 

It is easy to show that in the free falling EE system of Fig.\,\ref{Fig:FEwithEE} there is neither a Doppler nor a gravitational frequency shift. As ensured by (\ref{Eq:EEeq}) and assumption H4, the source and the detector maintain their original  zero velocities at all time, which excludes a Doppler shift. By the same equation, there is no gravitational field, hence there cannot be a gravitational redshift either.
Therefore, identical clocks tick the same regardless of their location in the Elevator, while photons travel at the same speed  $c\simeq3\cdot10^8\,\rm ms^{-1}$ as in FE, in agreement with special relativity.

The flawed analysis of AOK  is due to the fact that, while  they are themselves convinced to run the experiment in the free falling Einstein Elevator of  Fig.\,\ref{Fig:FEwithEE} --like the Observer $\rm{O_{EE}}$--   their analysis actually refers to (and it is correct for) a light source and a detector freely falling from different heights in the uniform gravitational field of the FE system as observed by $\rm{O_{FE}}$, who is rigid with it. %.

The proof is as follows, for both photons and clocks. 
Let us imagine that in the Flat Earth system  the free falling masses 
$m_1,m_2$ represent, respectively,  the light source and the detector as in the set up of AOK where the light signal travels against the gravitational acceleration $\vec{g}_{\oplus}$. The masses obey Eq.\,(\ref{Eq:FEeq}) --written by the Observer   $\rm{O_{FE}}$-- with assumption H4 (zero initial velocities), which ensures that they maintain their relative position and equal velocities at any given time, though positions and velocities change over  time. Since light is emitted by $m_1$ at time $t_{_E}$ while it is received by $m_2$ at a later time  $t_{_D}=t_{_E}+h/c$, at reception $m_2$ has acquired a velocity $g_\oplus h/c$ relative to that of $m_1$ at the time of emission (while moving towards it). The result is a positive Doppler shift (blueshift) $\Delta\nu_{_D}/\nu\simeq+g_\oplus h/c^2$.

At any given time the relative position of the two masses remains the same, and so does their separation distance along the gravity field lines. However, with a time difference $h/c$ between emission and detection there is a variation of the relative distance by $-\frac{1}{2}g_\oplus h^2/c^2$. As a result, while the first order gravitational shift (a redshift in this case)  $\Delta\nu_{grav}/\nu\simeq-g_\oplus h/c^2$ cancels exactly the Doppler blueshift due to the non zero relative velocity, there is an additional second order gravitational shift by $+\frac{1}{2}(g_\oplus h/c^2)^2$. Albeit extremely small, this term may appear as the net frequency shift to be expected for a light source and a detector freely falling from different heights in the Flat Earth system. However,  if in the previous calculation of the Doppler shift we take into account also the second order term, we find that it cancels the second order gravitational  contribution. 

A similar null result is expected also for clocks, and can be demonstrated as follows. 

Three identical clocks $C_\circ,C_1,C_2$ are proven to tick with identical periods $\tau_\circ=\tau_1=\tau_2$ when positioned at rest at the same height $h_\circ$ in the Flat Earth system of  Fig.\,\ref{Fig:FEsystem}. Then $C_1,C_2$ are moved to heights $h_1,h_2$ ($h_2=h_1+h$) and kept at rest\footnote{The rest position must obviously be guaranteed  by a support structure capable to provide the reaction force necessary to counteract the weight of the clock in the Flat Earth system.}. In their new locations the ticking periods $\tau'_1,\tau'_2$ of the clocks differ from the value measured at rest at the  initial altitude $h_\circ$ in the uniform gravitational field $g_{_\oplus}$ as $\tau'_1\simeq\tau_\circ\big(1-g_{_\oplus} h_1/c^2\big)$ and $\tau'_2\simeq\tau_\circ\big(1-g_{_\oplus} h_2/c^2\big)$ yielding a non zero shift relative to one another: 
\begin{equation}\label{Eq:shiftfreefall1}
\tau'_2-\tau'_1\simeq-\frac{g_{_\oplus}h}{c^2}\tau_\circ
\end{equation}
while still at rest at heights $h_1$ and $h_1+h$. Then, at a given time, both clocks are let free to fall with the acceleration $g_{_\oplus}$. $C_1$ passes by $C_\circ$, at  height $h_\circ$ with velocity $v_1$ and consequent time dilation $\tau''_1\simeq\tau_\circ\big (1+ \frac{1}{2}v_1^2/c^2\big)$. $C_2$ will pass by $C_\circ$ with  a higher velocity $v_2$, yielding a longer time dilation relative to it $\tau''_2\simeq\tau_\circ\big (1+ \frac{1}{2}v_2^2/c^2\big)$, with $v_2^2=v_1^2+2g_{_\oplus}h$. The result is an additional shift between the free falling clocks by:
\begin{equation}\label{Eq:shiftfreefall2}
\tau''_2-\tau''_1\simeq+\frac{g_{_\oplus}h}{c^2}\tau_\circ
\end{equation}
which nullifies the previous one, given by (\ref{Eq:shiftfreefall1}) acquired for having been placed at rest   at different altitudes  with the same separation distance $h$. 

In the context of this work it was mandatory to demonstrate with all details, step by step, that a radiation source and a detector, as well as two clocks, freely falling in the uniform gravitational field of the FE system from different heights do not undergo any frequency shift. However this result could be inferred from the fact that their frequencies must be shown to be identical to start with, while they are at rest at the same altitude in FE; then they are moved upwards and placed at rest at different altitudes from which they are finally let to fall freely, back to the same initial location, in the same reference system FE. In  so doing they simply go back to the same equal frequencies that  had been measured  at the initial  location, in the same gravitational field.

From these considerations we are led to conclude that the zero net effect we have demonstrated for photons up to the second order does in fact hold exactly (within assumptions H).

% 

% Please add the following required packages to your document preamble:
% \usepackage{graphicx}
\begin{table}[]
\caption{Frequency shifts for photons and clocks, gravitational (\textit{grav}) or Doppler (\textit{D}).\\ $1,2$:  lower and higher altitude  position, separation distance $h$, uniform gravitational acceleration $g_{_\oplus}$.\\  $S_1\rightarrow S_2$: photons emitted from $S_1$ towards $S_2$.\\ $\Delta\nu_{12}=\nu_2-\nu_1$; $\Delta\tau_{12}=\tau_2-\tau_1$ ($\tau=1/\nu$).}
\label{table:PhotonsAndClocks}
\resizebox{\columnwidth}{!}{%
\begin{tabular}{|l|l|l|}
\hline\hline
                              & \textsc{Photons}                                      &  \textsc{Clocks}                            \\ \hline \hline \hline
\tiny{1,2 in EE (Fig.\,\ref{Eq:FEeqElevator})}  & $\Delta\nu_{grav}=0$   & $\Delta\tau=0$                   \\ %\hline
                                                 & $\Delta\nu_{_D}=0$                           &                                         \\ \hline\hline
\tiny{1,2 at rest in PR  (Fig.\,\ref{Fig:PRandAS}\,(\textit{a})) } & $S_1\rightarrow S_2$                         & $C_1,C_2$           \\ %\hline
                           &             $\frac{(\Delta\nu_{12})_{grav}}{\nu}\simeq-\frac{g_{_\oplus}h}{c^2}$ &
  $\frac{\Delta\tau_{12}}{\tau}\simeq-\frac{g_{_\oplus}h}{c^2}$                                                           \\ %\hline
                              & $\Delta\nu_{_D}=0$                           &                                                               \\ \hline \hline
  \tiny{1,2 free falling in FE  (Fig.\,\ref{Fig:FEsystem})}  & $S_1\rightarrow S_2$                         & $C_1,C_2$                               \\ %\hline
      & $\frac{(\Delta\nu_{12})_{grav}}{\nu}\simeq$ &                                         \\ %\hline
 &
  $-\frac{g_{_\oplus}h}{c^2}+\frac{1}{2}\bigg(\frac{g_{_\oplus}h}{c^2}\bigg)^2$ &
  $\frac{\Delta\tau_{12}}{\tau}\simeq-\frac{g_{_\oplus}h}{c^2}$ \\ %\hline
                              & $\frac{(\Delta\nu_{12})_{D}}{\nu}\simeq$     &                                         \\ %\hline
 &
  $+\frac{g_{_\oplus}h}{c^2}-\frac{1}{2}\bigg(\frac{g_{_\oplus}h}{c^2}\bigg)^2$ &
  $\frac{\Delta\nu_{12}}{\nu}\simeq-\frac{g_{_\oplus}h}{c^2}$ \\ %\hline
                              & $\Rightarrow\ (\Delta\nu_{12})_{tot}=0$      & $\Rightarrow\ (\Delta\nu_{12})_{tot}=0$ \\ \hline\hline
\end{tabular}%
}
\end{table}

It is apparent that using a source and a detector, or two clocks, freely falling in a ground lab like the FE reference system is not the correct set up to measure the gravitational redshift because no frequency shift occurs in this case, neither for photons nor for clocks (as long as tidal effects can be neglected).  Indeed, the first measurement of the gravitational redshift by Pound and Rebka\,\cite{PoundRebka1960}, almost 50 years after its prediction, was performed with an emitter and an absorber of radiation at rest in the lab, certainly not in free fall. 

In Table\,\ref{table:PhotonsAndClocks} we summarize the frequency shifts --with their gravitational or Doppler physical origin-- for photons and clocks as discussed so far in the free falling Einstein Elevator  and in the Flat Earth reference system, in which case they can be either at rest (as in the Pound and Rebka, PR, reference system of Fig.\,\ref{Fig:PRandAS}(\textit{a}))  or free falling  in FE (as in Fig.\,\ref{Fig:FEsystem}).

The stated  aim  of AOK  was  to show that, as long as WEP holds,  there is no gravitational effect in a  reference system free falling in a uniform gravitational field (such as the Einstein Elevator of Fig.\,2). One wonders why they did not exploit the fact that the EE is equivalent to the S reference system in empty space of Fig. 3. With no gravity and no dynamics, they would have easily concluded, like the Observer $\rm{O_S}$, that there is no Doppler and no gravitational shift in this system, hence also in the equivalent EE system.

Indeed, the equivalence between the EE and the S systems is very helpful in answering an interesting and  subtle question. Consider an EE freely falling towards a flat, non rotating Moon, rather than the Earth, everything else being the same, in particular the assumptions H. From all what  $\rm{O_{EE}}$ can observe and measure inside his lab he cannot tell the difference. It is only by ``looking outside'' through the $z'=0$ plane, and by measuring the acceleration of the Elevator relative to the plane $z=0$ of the inertial system at rest on the surface of a flat, non rotating Moon that he will see the difference. 

Inside the free falling EE there is absence of weight but not absence of gravity. However, because of the equivalence between inertial and gravitational mass, Eq.\,(\ref{Eq:FEeqElevator2}) holds also as 
$\ddot z'=- g_{\leftmoon}+g_{\leftmoon}=0$, i.e. 
regardless of the fact that the local gravitational acceleration that cancels out  is  that of the Earth or of the Moon.
Hence, physics inside the free falling EE does not depend on the value of the acceleration of fall. 

A fact that  becomes obvious once the equivalence between the EE reference system (freely falling towards a flat, non rotating source body of arbitrary mass) and the reference system S (far away from any source mass) is established. It follows from this equivalence  that the actual value of the uniform acceleration of the gravity field does not affect the physics of either system.

This is no longer true globally. Since the gravity field is not uniform, geometry is no longer Euclidean and gravity is replaced by geometry whereby, as it is often summarized in a nice sentence by J. A. Wheeler: ``\textit{Matter tells spacetime how to curve;   spacetime tells test masses how to move}''. 

\section{Should the Pound and Rebka 1960 experiment be re-interpreted?}\label{Sec:PoundRebkaIsCorrect}

\noindent It took almost 50 years for Einstein's prediction\,\cite{Einstein1911} to be confirmed by the Pound and Rebka 1960 experiment\,\cite{PoundRebka1960}.  With a limited altitude difference in ground laboratories the term $g_\oplus h/c^2$ is small and the detection of gravitational redshift very challenging, more so in the presence of numerous disturbing effects.

In 1957 the first artificial satellite --Sputnik, Satellite in Russian--  was launched and the space era began. The first proposed space experiment was the measurement of the predicted gravitational frequency shift by launching a high precision $H$ maser clock in space (up to $10000\,\rm km$ altitude) with a rocket, and comparing it with an identical one on the ground. The advantage is the very large altitude difference compared to any tower on the ground. However, while in the lab the two clocks are both at rest, making the baseline of the experiment fixed, in this case the relative motion of the two clocks, due to the motion of the one in space and also of the one on the ground (because of  the motion of the Earth) requires to deal with various Doppler effects, competing directly with the signal of interest, which therefore must be reduced and/or modeled to the required level of accuracy. As the first high precision fundamental physics experiment in space the challenges were enormous. 
% Conclusions
Yet, it was the very significance of the experiment (promoted as a crucial test of General Relativity) to be immediately questioned by Leonard Schiff in 1960\,\cite{Schiff1960}, as we recall in the Conclusions.

By a remarkable coincidence, the M\"ossbauer effect was discovered in 1958\,\cite{Mossbauer1958}, worth the Nobel prize in 1961. It shows that a fraction of $\gamma$
rays are emitted by the nuclei of a solid without individual recoil of the nucleus itself, the recoil momentum being transferred to the crystal lattice as a whole. As a result the Doppler shift is greatly reduced.

\begin{figure}%[H]
\begin{center}
\includegraphics[width=0.20\textwidth]{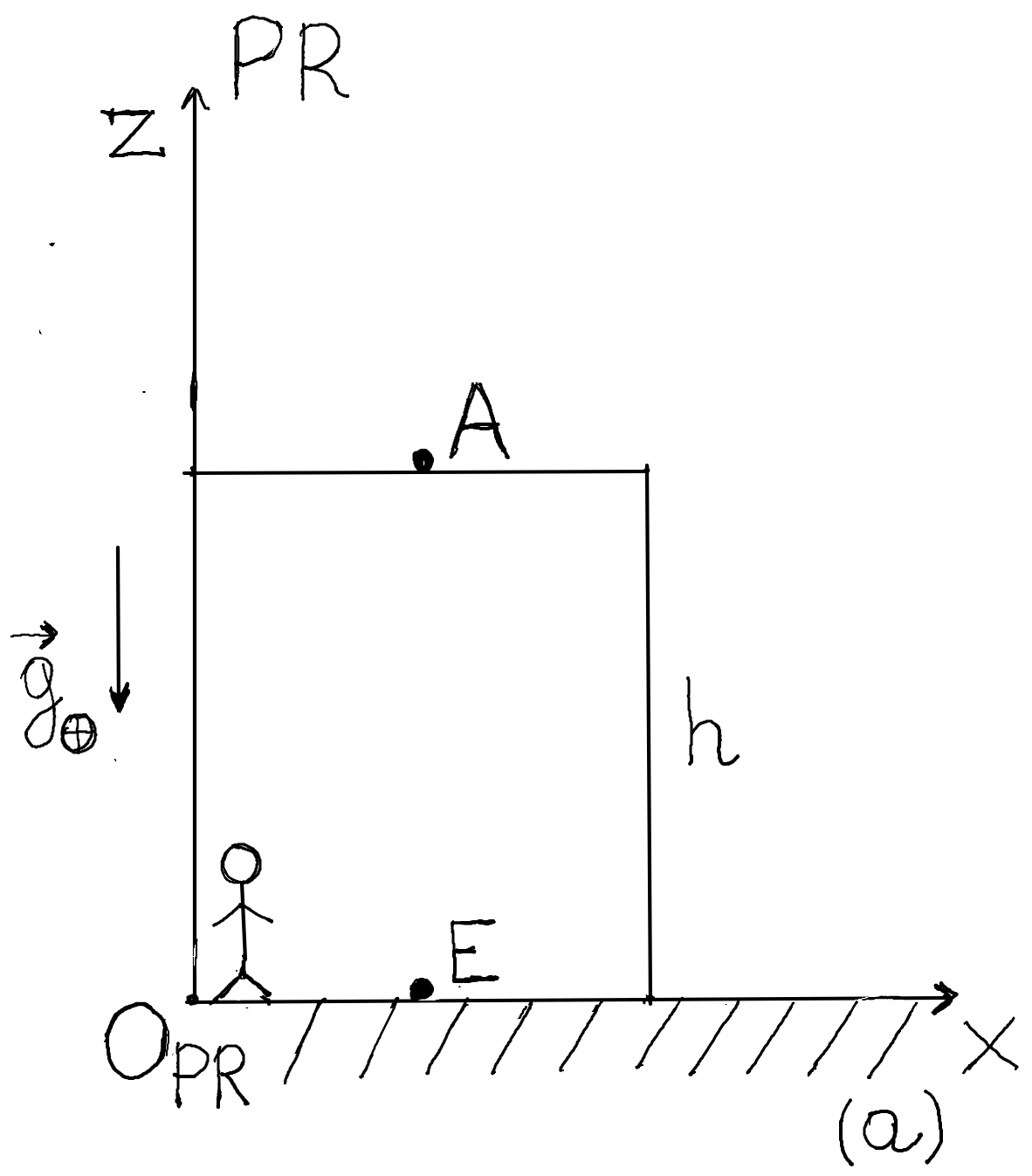}
\hspace{5mm}
\includegraphics[width=0.20\textwidth]{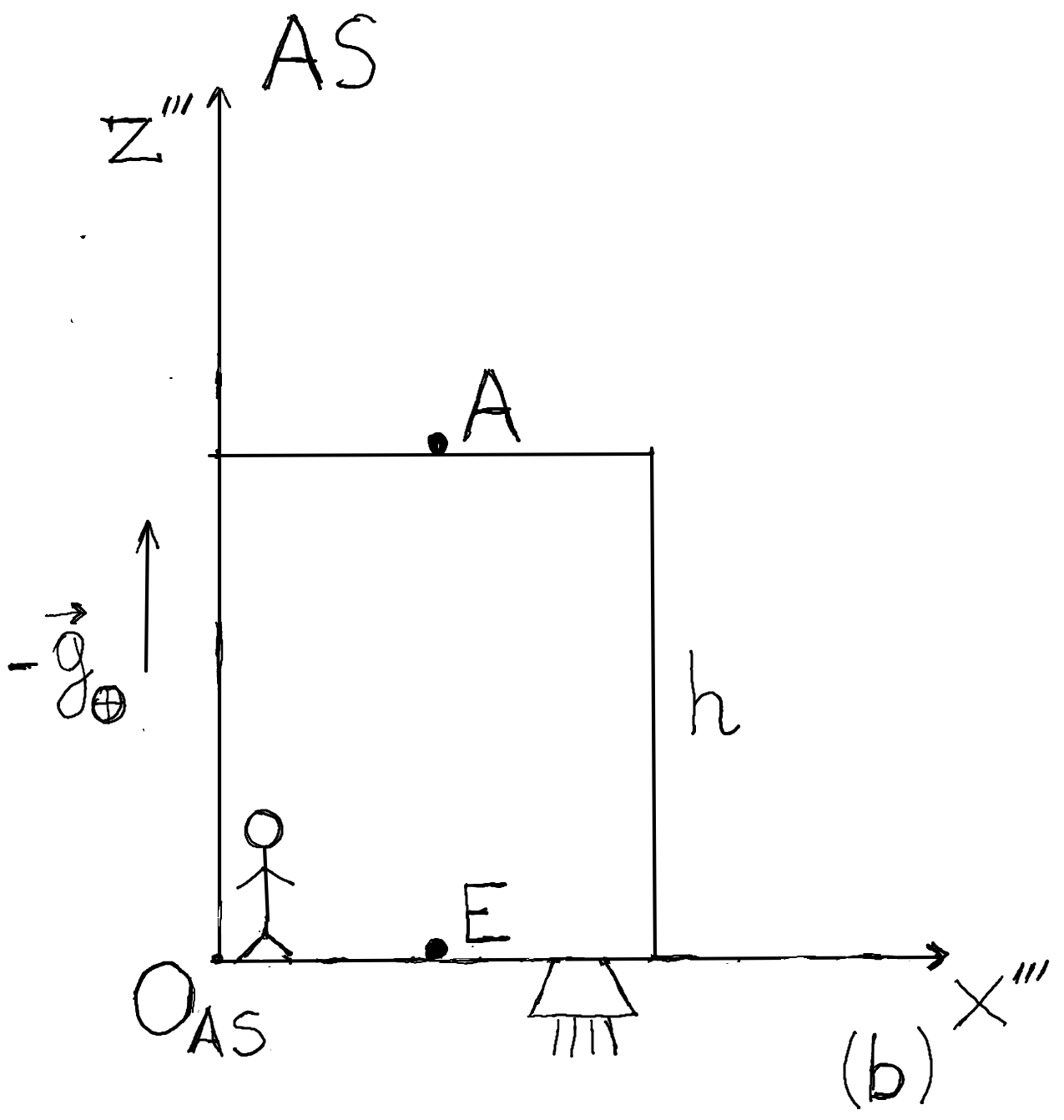}
\caption{(\textit{a}) Schematic representation in 2D of the reference system PR in which the Pound and Rebka experiment\,\cite{PoundRebka1960} was performed. It is the same as the inertial FE system of Fig.\,\ref{Fig:FEsystem} in which the  masses $m_1,m_2$ freely falling with $\vec{g}_\oplus$ are now replaced by the Emitter E and the Absorber A at rest in the lab. (\textit{b}) The reference system AS (Accelerated Spaceship) is attached to a ``spaceship'' equipped with a rocket capable of accelerating it with $-\vec{g}_\oplus$ far away from any source mass to ensure that there are no measurable gravitational effects in it. The axes $x''',z'''$ of the AS system are parallel to $x,z$ respectively. AS is  a non inertial accelerated system with an inertial acceleration field  $\vec{a}_i=+\vec{g}_\oplus$ (WEP holds) identical to the gravitational  field in PR. The Emitter and the Absorber are kept at rest by means of appropriate support structures as in PR. The reference systems PR and AS are equivalent.}
\label{Fig:PRandAS}
\end{center}
\end{figure}

This led Pound and Rebka in 1959\,\cite{PoundRebka1959} to propose a laboratory experiment to measure the gravitational frequency shift between $\gamma$ rays emitted and absorbed at different altitudes of an existing tower in the physics laboratories. As they pointed out,  the fixed baseline ``\textit{reduces unwanted Doppler shifts to only those resulting from thermal, seismic, or similar disturbances.}'' One year later\,\cite{PoundRebka1960} they published the results of the first experimental measurement of the gravitational shift as the ratio between the frequency shift measured in the experiment and its theoretical value:
\begin{equation}\label{Eq:PoundRebkaResult1960}
\frac{(\Delta\nu)_{exp}}{(\Delta\nu)_{th}}=+1.05\pm0.10
\end{equation}
soon improved by one order of magnitude to $1\,\%$ accuracy\,\cite{PoundSnider1960}.

Twenty years later, in 1980, the result  of the experiment with one $H$ maser clock on the ground and an identical one in space (later known as the GP-A or Gravity Probe A mission) reported an agreement between the observed and the theoretical gravitational shift to $0.007\%$, an improvement by more than two orders of magnitude\,\cite{Vessot1980}. 

The reference system of the Pound and Rebka experiment (the PR system) is sketched in Fig.\,\ref{Fig:PRandAS}(\textit{a}). It is the same as the FE system of Fig.\,\ref{Fig:FEsystem} and it is an inertial reference system. The only difference concerns the masses $m_1,m_2$ representing the emitter and the absorber which are now rigidly attached to the PR system (indicated as E and A in Fig.\,\ref{Fig:PRandAS}(\textit{a})) and not in free fall like $m_1,m_2$ in FE. The Observer  $\rm{O_{PR}}$  argues that between the emitter and the receiver there is a gravitational potential difference $g_{_\oplus} h$, and therefore there must be a gravitational frequency shift. He also argues that, as long as  the Doppler disturbances mentioned in\,\cite{PoundRebka1959} are taken care of, there is no major Doppler shift to compete with the gravitational one.
 
 The PR and AS systems, as depicted in Fig.\,\ref{Fig:PRandAS} and described in the caption, are equivalent to each other. The Observers $\rm{O_{PR}}$ and $\rm{O_{AS}}$ write the same equations of motion and predict a gravitational redshift only, with no Doppler shift to cancel it out.

In Figure\,2(b) of Ref.\,\cite{AOK2024} AOK sketch the emitter and the absorber of $\gamma$ rays used in the Pound and Rebka experiment --at rest in the lab at different heights-- as subjected to the local gravitational acceleration of the Earth plus an equal and opposite non-gravitational acceleration, which they name $a_{_{NG}}$, to counteract their weight and keep them at rest as in Fig.\,\ref{Fig:PRandAS}(\textit{a}).\footnote{In the rebuttal   \href{https://doi.org/10.1088/1402-4896/aeb33a}{(AOK, 2026)} the Authors point out that in fact they do not assume   $a_{_{NG}}$ to equal $g_{_\oplus}$  (in modulus). However, since they conclude that Pound and Rebka have measured a Doppler shift due to  $a_{_{NG}}$, it follows that within their analysis   the two must have the same modulus. Moreover,  $a_{_{NG}}$ must be the same for both the light source and the detector.} 
Then, they analyze the effect of the gravitational acceleration separately from that of $a_{_{NG}}$. They argue that in the absence of  $a_{_{NG}}$ there would be a gravitational and a Doppler shift cancelling each other out, as in the case they had previously analyzed.\footnote{Ref.\,\cite{AOK2024} describes in Figure\,2(a)  an emission/detection experiment carried out in the free falling EE while in fact their analysis refers to the case of a source and a detector free falling in the FE  system in which, indeed, the gravitational and Doppler shifts cancel each other out, as shown in Sec.\,\ref{Sec:Figure2aEE}.} Then they deduce that the non-gravitational acceleration $a_{_{NG}}$ gives rise to a non zero relative velocity $a_{_{NG}}h/c$ between the absorber at the time of reception and the emitter at the time of emission $h/c$ seconds earlier (moving them farther apart), hence to a Doppler relative frequency redshift by $-|a_{_{NG}}|h/c^2$. Thus they state that this is the redshift that  Pound and Rebka have actually measured, not the gravitational redshift predicted by Einstein in 1911.

Should this be the case, the physical relevance of their experiment, so far regarded as the first experimental evidence of gravitational redshift, would be highly diminished. 

We refute AOK's argument as follows.  The emitter and the absorber, with masses $m_1,m_2$ are at rest in the PR system as shown in Fig.\,\ref{Fig:PRandAS}\textit{(a}) because the structure underneath each of them provides the non-gravitational reaction forces  $R_1$, $R_2$ in the $+z$ direction necessary to balance their weight.\footnote{The reaction force is calculated from the applied gravitational force (the weight) of each body. It is therefore the  forces that matter, not  the accelerations. To make this fact absolutely clear we explicitly use here the reaction forces $R_1$, $R_2$, unlike in  \href{https://doi.org/10.1088/1402-4896/aeb1fd}{(Nobili and Anselmi, 2026)}}
The reaction force is the macroscopic representation of the microscopic forces acting at the points of contact of the body with the supporting structure, it is perpendicular to it, depends on the weight of the body to be supported, i.e. on its gravitational mass  --up to the structural limit of the support-- and disappears if there is no contact.

Reaction forces are punctual and temporary. They do not generate a non-gravitational field $\vec{a}_{_{NG}}$ as envisaged by AOK  in addition to the gravitational field $\vec{g}_{_\oplus}$ (which is \textit{always} present \textit{at any point} in the lab, whether a mass is located there or not). Therefore reaction forces  cannot be treated as existing independently from the emitter and the absorber being at rest at their individual locations in the lab.  If they are still, there is neither a velocity nor a relative velocity, hence there can be no Doppler shift from such forces. On the other hand, if the masses are not supported by the structure there is no reaction force, hence no Doppler shift can be generated by a non-existing force. By Occam's razor, the issue is set.  However, it is worth adding the following general consideration.

Should the Pound and Rebka experimental set up be equivalent to the one envisaged in Figure\,2(b) of Ref.\,\cite{AOK2024}, the emitter and the absorber would be in free fall with $\vec{g}_{_\oplus}$ unless there is an acceleration equal and opposite to it, anywhere in the lab, to prevent them from falling. The force required should be non-gravitational, and such that it produces the same acceleration on both the (otherwise) free falling emitter and absorber.  As an example, consider a source of radiation capable to produce a force antiparallel to the local gravitational force anywhere in the lab; the only way for it to compensate the acceleration of gravity with which emitter and absorber are falling is for them to have the same area-to-mass ratio. This is well known for artificial satellites  subjected to solar radiation pressure, or atmospheric drag from residual air density or many other non-gravitational forces as well (see e.g.\,\cite{Librino1987}). Gravitation is the only force producing the same acceleration on all bodies regardless of their mass and composition and no experiment so far as proved otherwise.

In the  Pound and Rebka experiment with emitter and absorber rigidly attached to the lab there is no non-gravitational field equal and opposite to the gravitational one (as erroneously assumed by AOK), the set up is the same as envisaged in Einstein\,\cite{Einstein1911}, and they measure a frequency shift of gravitational origin as predicted therein.

Having concluded that Pound and Rebka did instead measure a Doppler shift due to the non-gravitational acceleration  ${a}_{_{NG}}$, AOK announce that ``\textit{A local redshift test in free fall, where  ${a}_{_{NG}}=0$, should be possible in the near future.}'' With ${a}_{_{NG}}=0$, hence with no supporting structure, the emitter and the absorber would be free falling in the lab like the masses $m_1,m_2$ in the Flat Earth system of Fig.\,\ref{Fig:FEsystem}, observed by  $\rm{O_{FE}}$. We have shown in Sec.\,\ref{Sec:Figure2aEE} that in this case the gravitational and Doppler shift cancel out for both photons and clocks. This is the reason --in addition to many practical  ones--  why  Pound and Rebka did put emitter and absorber at rest in the lab, and never suggested to let them fall freely. Should AOK manage to realize their planned experiment, and to perform it correctly, they will observe no frequency shift, not even to second order.
Should they find a non zero shift, it may be due to a deviation from the assumed uniform gravity field, i.e. it may be a tidal effect.

\section{Conclusions}\label{Sec:Conclusions}   

\noindent This paper was prompted by the bold statement by AOK at the start of Ref.\,\cite{AOK2024} that ``\textit{The influence of a uniform gravitational field (UGF) on matter waves and clocks is described inconsistently throughout research and education"}. A UGF is a convenient approximation widely adopted in introductory texts but its use is not without perils, and coming back to it to get rid of any inconsistencies is always worthwhile. Unfortunately, we find Ref.\,\cite{AOK2024} itself presents an inconsistent picture of the gravitational redshift in a UGF, which leads the authors to the extraordinary statement that the foundational Pound and Rebka experiment\,\cite{PoundRebka1960} did not observe – for the first time – a gravitational redshift but ``\textit{the Doppler shift associated with the non-gravitational acceleration}'' due to the reaction force of the supports.

The flawed analysis of AOK  goes back to a mix-up of a Flat Earth reference system and a free-falling Einstein Elevator system (Sec.\,\ref{Sec:Figure2aEE}).  The analysis of a  light source and  a detector freely falling in the UGF of a Flat Earth system, as seen by an observer rigidly fixed in it, is attributed to an observer free falling with them. To prove our case, we proceed through the definition of the reference systems involved and the equations of motion applying to each, from which the correct  results  follow trivially once the right reference system is adopted. As a consequence, our analysis restores the famous 1960 experiment by Pound and Rebka to its rightful place in the history of Physics (Sec.\,\ref{Sec:PoundRebkaIsCorrect}). 
 
As  remarked by Leonard Schiff\,\cite{Schiff1960}, measuring the gravitational redshift does not test General Relativity since it requires only the WEP --on which General Relativity is founded-- and special relativity. Direct measurements have proceeded from the pioneering experiments of Pound, Rebka and Snider\,\cite{PoundRebka1959,PoundRebka1960,PoundSnider1960}, to the first fundamental physics experiment in space\,\cite{Vessot1980}, to recent  improved versions of it\,\cite{GalileosGravRed1,GalileosGravRed2}.
In a timespan of about 60 years the accuracy has improved from $10^{-2}$\,\cite{PoundSnider1960}  to about $7\cdot10^{-5}$\,\cite{Vessot1980} to about $2\cdot10^{-5}$\,\cite{GalileosGravRed1,GalileosGravRed2}, the latter result seeming very hard to improve, owing to the difficulty of keeping under control the systematic errors over the very long baselines that are needed to generate a strong signal.   

Following  Schiff\,\cite{Schiff1960}, measurements of the gravitational redshift cannot go beyond providing corroborative evidence to the available tests of the WEP (as long as WEP holds for clocks as well as for ordinary matter, see\,\cite{Ashby2007}). Whereas gravitational redshift experiments are absolute measurements of a physical quantity that must be   accurately predicted and recovered from the data, WEP tests can be designed and performed as null experiments, which are known to be the most precise and accurate experiments in physics. Hence  their much deeper probing power, as quantified in\,\cite{AJP2013} in 2013. Since then, the first WEP test in space has reported a further 
improvement by about 2 orders of magnitude, to $\simeq10^{-15}$\,\cite{MicroscopeFinalResults}. Although issues have been raised and ambiguities have been identified, suggesting an alternative approach to the data analysis of that mission\,\cite{OnMicroscopeFinalResults}, it is clearly impossible for measurements of the gravitational redshift to bridge the gap with tests of the WEP. 

Finally, a consideration  is due on the rebuttal \href{https://doi.org/10.1088/1402-4896/aeb33a}{(AOK, 2026)}.\footnote{The Authors maintain that the Pound and Rebka experiment has measured the frequency shift due to the same non-gravitational acceleration  $a_{_{NG}}$ acting on both  the emitter and the absorber. Non-gravitational forces do not obey the equivalence principle, hence they cannot produce the same acceleration on  bodies of different mass, as in this case. It is therefore  impossible to define a unique  $a_{_{NG}}$  for both of them.  Yet, the rebuttal concludes that ``\textit{The Pound-Rebka experiment was the first test of the equivalence principle in a relativistic setting and is therefore rightfully celebrated.}''  How can a non-gravitational force, which does not obey the equivalence principle,   play a crucial rôle in testing the equivalence principle is beyond our comprehension. } 

\vspace{0.5cm}
\noindent \textbf{Acknowledgments} Thanks are due to Eric Adelberger for his insightful comments.
 
%\newpage

%\bibliographystyle{apsrev4-2}
\bibliography{prova.bib}
\end{document}